\documentclass[preprint,showpacs,preprintnumbers,amsmath,amssymb,superscriptaddress,epsfig]{article}

\usepackage{epsfig}

\begin{document}

\begin{center}
{\Large 
{\bf The Supersymmetric $(B-L)$ model with three non-identical right-handed neutrinos}}\\
M. C. Rodriguez \\
{\it Departamento de F\'\i sica \\
Universidade Federal Rural do Rio de Janeiro - UFRRJ \\
BR 465 Km 7, 23890-000, Serop\'edica, RJ, Brazil, \\
email: marcoscrodriguez@ufrrj.br \\
}
\end{center}

\begin{abstract}
We build a supersymmetric version with 
$SU(3)_{C}\otimes SU(2)_{L}\otimes U(1)_{Y^{\prime}}\otimes U(1)_{(B-L)}$ 
gauge symmetry, where $Y^{\prime}$ is a new charge and ($B$) and ($L$) 
are the usual baryonic and leptonic numbers, respectivelly.
The model has three right-handed neutrinos with non identical $(B-L)$ charges. We will use the superfield formalism in order to build our lagrangian and it is possible to accommodate all fermion masses at the tree level. In particular, the type-I seesaw mechanism is implemented for the generation of the active neutrino masses and also explain the mixing angle at the fermion sector. There are good candidates 
to Dark Matter, it can be the lighest righ-handed 
neutrinos or lighest right-handed sneutrinos or the lighest usual 
scalar field and due a Majorana phase at sneutrinos right-handed masses, it can induce Leptogenesis in this model.
\end{abstract}
PACS number(s): 12.60.Cn; 12.60. Jv; 14.60.Pq; 14.60.St.

\section{Introduction}
\label{sec:intro}

Although the Standard Model (SM) gives very good results, there are 
two evidences wchich hint the physics beyond the SM
\begin{itemize}
\item The recent groundbreaking discovery of nonzero neutrino masses and
oscillations;
\item The strength of CP violation in the SM it is not possible to generate sufficient baryon 
asymmetry through the phase of the Cabibbo-Kobayashi-Maskawa (CKM) mixing matrix.
\end{itemize}

The generation of neutrino masses is an important issue in any realistic 
extension of the SM and in general, the values of these masses that are needed to explain all neutrino 
oscillation data are not enough to put strong constraints on model building. It means that several models can induce neutrino
masses and mixing compatible with experimental data \cite{Rodriguez:2016esw,Rodriguez:2020}. So, instead of 
proposing models built just to explain the neutrino properties, it is more
useful to consider what are the neutrino masses that are predicted in any
particular model which has motivation other than the explanation of neutrino
physics.

However in the Minimal Supersymmetric Standard Model (MSSM) contain new CP-violating
sources beyond the CKM matrix \cite{Rodriguez:2016esw,Rodriguez:2020,Rodriguez:2019,drees,tata,
Khalil:2009tm,Kajiyama:2009ae}. By another hand Leptogenesis  \cite{yanagida} is based on a high scale seesaw mechanism, provides an
attractive scenario to explain the baryon asymmetry and in
this scenario, supersymmetry should be introduced to stabilize the
electroweak scale. Therefore, Leptogenesis is more natural in supersymmetric
models \cite{Khalil:2009tm,Kajiyama:2009ae}. 

Recently, a new Leptogenesis scenario, soft leptogenesis, 
has been proposed, where sneutrino decays offer new possibilities for generating the matter asymmetry \cite{grossman,D'Ambrosio:2003wy,Grossman:2005yi}. We can 
get the following superpotential
\begin{eqnarray}
W=W^{MSSM}+
(Y_{\nu})_{ij}\left( \hat{L}_{i}\hat{H}_{2}\right) \hat{N}_{j}+
\frac{1}{2}(Y_{N})_{ij}\hat{N}_{i}\hat{N}_{j}\hat{\chi}_{1}+
\mu^{\prime}\hat{\chi}_{1}\hat{\chi}_{2},
\end{eqnarray}
where $W^{MSSM}$ is the superpotential defined at MSSM. The scenarios presented have assume universal soft SUSY \cite{Khalil:2009tm,Kajiyama:2009ae}
\begin{equation}
{\cal L}_{\rm soft} = \frac{\tilde m_{N}^{2}}{2} \tilde{N}^{\dag}
\tilde{N}+ \frac{B_{M}^{2}}{2} \tilde{N} \tilde{N} +
B \mu^{\prime}\chi_{1}\chi_{2} + 
A_{\nu}Y_{\nu} \left( \tilde{L} H_{2} \right) \tilde{N} +
A_{N}Y_{N}NN \chi_{1}+ hc, 
\label{egipciobl}
\end{equation}
and in the case of mSUGRA, $B_{M}^{2}\equiv B_{N} M_{N}$ and 
$M_{N}=Y_{N} \langle \chi_{1} \rangle$. This sector has one physical CP violating phase 
\begin{equation}
\phi_\nu = {\rm arg}(A_{\nu} B^{*}_{N}) , 
\end{equation}
therefore there are a mixing between the sneutrino $\tilde{N}$ and
the anti-sneutrino $\tilde{N}^{\dag}$. The CP violation in the $\tilde{N}$-mixing is induced by 
the phase $\phi_\nu$ generates lepton asymmetry in the final states of the
$\tilde{N}$-decay. This lepton asymmetry is converted to baryon
asymmetry through the sphaleron process \cite{spha}. 

Recently, it was proposed some interestings models with $(B-L)$ gauge 
symmetry where the new gauge symmetry is defined as
$SU(3)_{C}\otimes SU(2)_{L}\otimes U(1)_{Y^{\prime}}\otimes U(1)_{(B-L)}$. In this model, the parameter $Y^{\prime}$ is
chosen to obtain the hypercharge $Y$ of the Standard Model (SM), given by
$Y=Y^\prime+(B-L)$, together with the followings assumptions \cite{Montero:2007cd}:
\begin{enumerate}
\item Only left-handed neutrinos are the active neutrinos;
\item ($L$) assignment is restricted to the integer numbers;
\end{enumerate}
the anomaly cancellation imply only three right-handed
neutrinos can be added to the minimal representation of SM and 
arise two types of model \cite{Montero:2007cd}:
\begin{enumerate}
\item The assignment $(B-L)=-1$ for all sterile neutrinos;
\item Two of the sterile neutrinos have $(B-L)=4$ and the third one \newline
$(B-L)=-5$.
\end{enumerate} 
There is an interesting model having fractional $(B-L)$ charges \cite{Bernal:2018aon}.   

The supersymmetric version of this model with three identical neutrinos was presented 
in Ref.~\cite{Montero:2016qpx}. In this article we are going to present the supersymmetrics 
version for this kind of model with three non identical neutrinos 
where two right-handed neutrinos having $(B-L)=-4$ and the third one having $(B-L)=5$ and we can get similar results for leptogenesis as 
presented at \cite{Khalil:2009tm,Kajiyama:2009ae}.

The outline of this paper is as follows: In Sec.(\ref{sec:model}) we present the model, in Sec.(\ref{Rparitym2}) we define a $R$-Parity 
in such way that neutrinos and neutralinos are distinguish particles. At Sec.(\ref{sec:lagrangianm2}) we present the 
lagrangian of this model in the superfield formalism and after it we calculate the masses to the all usual leptons. We present some results 
about Leptogenesis at Sec.(\ref{sec:leptom1}) and at the end we 
present our conclusions.   

\section{The Model}
\label{sec:model}
 
The gauge symmetry in this model is 
\begin{equation}
SU(3)_{C}\otimes SU(2)_{L}\otimes U(1)_{Y^{\prime}}\otimes U(1)_{(B-L)},
\end{equation} 
and the charge operator is given by
\begin{equation}
\frac{Q}{e}=I_{3}+ \frac{1}{2} \left[ Y^{\prime}+ \left( B-L \right) \right] .
\end{equation}

We start introducing the leptons of our model, as usual in supersymmetric models, in the following chiral superfields  ($i=1,2,3$, is a familly indices):
\begin{eqnarray}
&&\hat{L}_{iL}= \left(
\begin{array}{c}
\hat{\nu}_{i} \\
\hat{l}_{i}
\end{array}
\right)_{L}  \sim (1,{\bf 2},0,-1),\;\;
\hat{E}_{iR} \sim (1,{\bf 1},1,1), \nonumber \\
\end{eqnarray}
in parenthesis it appears the transformations properties under the 
respective factors $(SU(3)_{C},SU(2)_{L},U(1)_{Y^{\prime}},U(1)_{(B-L)})$. We 
will not discuss the quarks because, they are the same as presented 
at \cite{Montero:2007cd,Montero:2016qpx,Machado:2010ui}. 

There are, also, three right-handed  neutrinos
\begin{eqnarray}
&&
\hat{N}_{1R} \sim (1,{\bf 1},5,-5), \,\
\hat{N}_{\beta R} \sim (1,{\bf 1},-4,4) , \nonumber \\
\end{eqnarray}
where $\beta =2,3$.

The Higgs sector of this model, we have the usual doublet $\hat{H}_{1,2}$
\begin{equation}
\hat{H}_{1} = \left(
\begin{array}{c}
\hat{h}^{+}_{1} \\
\hat{h}^{0}_{1}
\end{array}
\right) \sim (1,{\bf 2},1,0), \,\
\hat{H}_{2} = \left(
\begin{array}{c}
\hat{h}^{0}_{2} \\
\hat{h}^{-}_{2}
\end{array}
\right) \sim (1,{\bf 2},-1,0),
\label{3t}
\end{equation}
while their vev as usual are given by:
\begin{equation}
\langle H_{1} \rangle = \frac{v_{1}}{\sqrt{2}}, \,\ 
\langle H_{2} \rangle = \frac{v_{2}}{\sqrt{2}}.
\label{vevh1h2}
\end{equation}

It is necessary to enlarge the scalar doublet sector adding four new scalars 
\begin{eqnarray}
\hat{\Phi}_{1} &=& \left(
\begin{array}{c}
\hat{\phi}^{0}_{1} \\
\hat{\phi}^{-}_{1}
\end{array}
\right) \sim (1,{\bf 2},5,-6), \,\
\hat{\Phi}^{\prime}_{1} = \left(
\begin{array}{c}
\hat{\phi}^{\prime +}_{1} \\
\hat{\phi}^{\prime 0}_{1}
\end{array}
\right) \sim (1,{\bf 2},-5,6), \nonumber \\
\hat{\Phi}_{2} &=& \left(
\begin{array}{c}
\hat{\phi}^{0}_{2} \\
\hat{\phi}^{-}_{2}
\end{array}
\right) \sim (1,{\bf 2},-4,3), \,\
\hat{\Phi}^{\prime}_{2} = \left(
\begin{array}{c}
\hat{\phi}^{\prime +}_{2} \\
\hat{\phi}^{\prime 0}_{2}
\end{array}
\right) \sim (1,{\bf 2},4,-3).
\label{extra3t}
\end{eqnarray}
The vevs of the new scalars can be written in the following form:
\begin{eqnarray}
\langle \Phi_{1}\rangle &=& \frac{u_{1}}{\sqrt{2}},  
\langle \Phi^{\prime}_{1}\rangle =
\frac{u^{\prime}_{1}}{\sqrt{2}},  
\langle \Phi_{2}\rangle = \frac{u_{2}}{\sqrt{2}}, 
\langle \Phi^{\prime}_{2}\rangle = \frac{u^{\prime}_{2}}{\sqrt{2}}. 
\label{vevnewdoublets}
\end{eqnarray}

In order to obtain an arbitrary mass matrix for the neutrinos we have to introduce the following additional singlet
\begin{eqnarray}
\hat{\varphi}_{1} \sim (1,{\bf 1},8,-8),  \,\
\hat{\varphi}_{2} \sim (1,{\bf 1},-10,10).
\label{singletsextra}
\end{eqnarray}
The vevs of the new scalars can be written in the following form:
\begin{eqnarray}
\langle \varphi_{1}\rangle &=& \frac{w_{1}}{\sqrt{2}},
\langle \varphi_{2}\rangle = \frac{w_{2}}{\sqrt{2}}. 
\label{vevnewsinglets}
\end{eqnarray}

We could introduce
\begin{eqnarray}
\hat{\varphi}_{3} \sim (1,{\bf 1},-1,1),  
\label{singletsextraforbidd}
\end{eqnarray}
and this new scalar would induce mixing in the right handed neutrinos sector, we can 
avoid this mixing term with $R$-Parity, as we will discuss at Sec.(\ref{Rparitym2}) and Eq.(\ref{mixingneutrinos}).

Concerning the gauge bosons and their superpartners, they are introduced in vector superfields. See Table~\ref{table3} the particle content together with the gauge coupling constant of each group.
\begin{table}[h]
\begin{center}
\begin{tabular}{|c|c|c|c|c|c|}
\hline
${\rm{Group}}$ & ${\rm Vector}$ & ${\rm{Gauge}}$ & ${\rm{Gaugino}}$ & ${\rm{Auxiliar}}$ & ${\rm constant}$ \\
\hline
$SU(2)_{L}$ & $\hat{W}^{i}$ & $W^{i}$ & $\tilde{W}^{i}$ & $D_{W}$ & $g$ \\
\hline
$U(1)_{Y^{\prime}}$ & $\hat{b}_{Y^{\prime}}$ & $b_{Y^{\prime}}$ & $\tilde{b}_{Y^{\prime}}$ & $D_{Y^{\prime}}$ & $g_{Y^{\prime}}$ \\
\hline
$U(1)_{B-L}$ & $\hat{b}_{BL}$ & $b_{BL}$ & $\tilde{b}_{BL}$ & $D_{BL}$ & $g_{BL}$ \\
\hline
\end{tabular}
\end{center}
\caption{Information on fields contents of each vector superfield of this model and their gauge constant.}
\label{table3}
\end{table}

These are the minimal fields, we need to construct this supersymmetric model.

\section{R-Parity}
\label{Rparitym2}

Let us begining defining the $R$-parity in the model with the particle content listed above. We define at Tab.(\ref{rcharge}) the $R$-charge 
($n_{\Phi}$)\footnote{Here $\Phi$ means chiral superfield as defined at \cite{Rodriguez:2016esw,Rodriguez:2020,Rodriguez:2019,drees,tata}.} of each superfield in our model. Using these 
$R$-charges we can get the $R$-Parity of each fermion field contained in these chiral superfield these results we shown at Tab.(\ref{fermionblrm2}).
\begin{table}[h]
\begin{center}
\begin{tabular}{|c|c|c|c|c|c|}
\hline 
${\rm{Superfield}}$ & $\hat{L}_{iL}$ & $\hat{E}_{iR}$  & $\hat{N}_{1R}$ & $\hat{N}_{\beta R}$ & $\hat{S}$  \\
\hline 
$R-{\rm{charge}}$ & $n_{L}=1$ & $n_{E}=-1$ & $n_{N_{1}}=-1$ & $n_{N_{\beta}}=-1$ & $n_{S}=0$    \\
\hline
\end{tabular}
\end{center}
\caption{Information about the $R$-charge ($n_{\Phi}$) of all the chiral superfields of this model, our notation here 
$S=H_{1,2}, \Phi_{1,2}, \Phi^{\prime}_{1,2}, \varphi_{1,2}$.}
\label{rcharge}
\end{table}
\begin{table}[h]
\begin{center}
\begin{tabular}{|c|c|c|c|c|c|c|c|c|}
\hline 
$(B-L)$ & $-5$ & $4$ & $-6$ & $6$ & $3$ & $-3$ & $-8$ & $10$     \\
\hline 
${\rm{Fermion}} $ & $N_{1R}$ & $N_{\beta R}$ & $\tilde{\Phi}_{1}$ & $\tilde{\Phi}^{\prime}_{1}$ & $\tilde{\Phi}_{2}$ & $\tilde{\Phi}^{\prime}_{2}$ & 
$\tilde{\varphi}_{1}$ & $\tilde{\varphi}_{2}$     \\
\hline 
$R-{\rm{Parity}}$ & $+1$ & $+1$ & $-1$ & $-1$ & $-1$ & $-1$ & $-1$ & $-1$     \\
\hline
${\rm{Scalar}} $ & $\tilde{N}_{1R}$ & $\tilde{N}_{\beta R}$ & $\Phi_{1}$ & $\Phi^{\prime}_{1}$ & $\Phi_{2}$ & $\Phi^{\prime}_{2}$ & 
$\varphi_{1}$ & $\varphi_{2}$    \\
\hline 
$R-{\rm{Parity}}$ & $-1$ & $-1$ & $+1$ & $+1$ & $+1$ & $+1$ & $+1$ & $+1$     \\
\hline
\end{tabular}
\end{center}
\caption{Information about the $(B-L)$ quantum number and $R$-Parity of new fields, fermions and scalars, of this model.}
\label{fermionblrm2}
\end{table}

\section{Full Lagrangian}
\label{sec:lagrangianm2}

With the superfields presented at Sec.(\ref{sec:model}), we can built a
supersymmetric invariant lagrangian in the 
superfield formalism \cite{Rodriguez:2016esw,Rodriguez:2020,Rodriguez:2019,drees,tata}. It has the following form
\begin{equation}
{\cal L} = {\cal L}_{SUSY} + {\cal L}_{soft}. \label{l1}
\end{equation}
Here, as usual, ${\cal L}_{SUSY}$ is the supersymmetric piece, see Sec.(\ref{subsec:stm2}), while ${\cal L}_{soft}$ explicitly breaks SUSY, see Sec.(\ref{sec:softterms}).

\subsection{Supersymmetric Piece}
\label{subsec:stm2}

We can rewrite each term appearing in Eq.(\ref{l1}), in the following way
\begin{equation}
{\cal L}_{SUSY} = {\cal L}_{lepton} + {\cal L}_{scalar} + {\cal L}_{gauge} + {\cal L}_{quark},
\label{l2}
\end{equation}
the last two terms above, ${\cal L}_{gauge}, {\cal L}_{quark}$, are the same as presented at the supersymmetric model with 
three identical neutrinos presented in \cite{Montero:2016qpx}.

\begin{figure}[ht]
\begin{center}
\vglue -0.009cm
\mbox{\epsfig{file=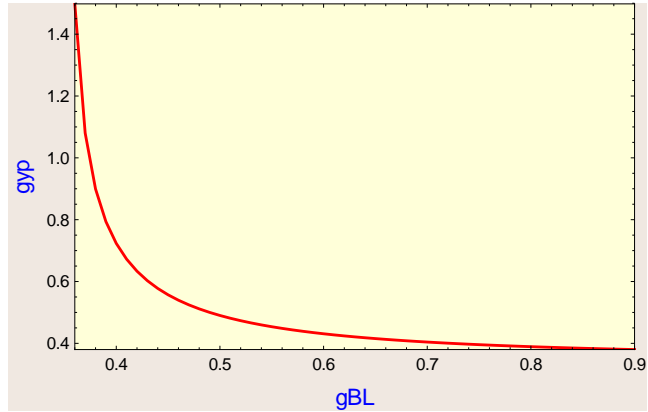,width=0.7\textwidth,angle=0}}
\end{center}
\caption{The numerical values for $g_{Y^{\prime}}$ in terms of 
$g_{BL}$, those values were obtained using Eq.(\ref{runcouplingconstant}).}
\label{fig0}
\end{figure}

The first term, ${\cal L}_{lepton}$, written at Eq.(\ref{l2}) is given by:
\begin{eqnarray}
{\cal L}_{lepton}&=&\left[ \int d^{4}\theta\;\left( 
{\cal L}^{char}+{\cal L}^{N_{1}}+{\cal L}^{N_{\beta}} \right) \right],
\label{leptontotal}
\end{eqnarray}
where we have defined
\begin{eqnarray} 
{\cal L}^{char}&=& \sum_{i=1}^{3}\left[\,
\hat{ \bar{L}}_{iL}e^{2[g\hat{W}+g_{BL} \left( -\frac{1}{2} \right) \hat{b}_{BL}]} \hat{L}_{iL}+
\hat{ \bar{E}}_{iR} e^{2[g_{Y^{\prime}} \left( \frac{1}{2} \right)
	\hat{b}_{Y^{\prime}}+g_{BL} \left( \frac{1}{2} \right) \hat{b}_{BL}]} \hat{E}_{iR} \right], \nonumber \\ 
\label{chargedlepton} 
\end{eqnarray}
In the expressions above we have used $\hat{W}=(T^{a}\hat{W}^{a})$ where $T^{a}=( \sigma^{a}/2)$ (with $a=1,2,3$) are the
generators of $SU(2)_{L}$ while $g_{Y^{\prime}}$ and $g_{BL}$ are the gauge constant constants of the $U(1)_{Y^{\prime}}$ and the
$U(1)_{(B-L)}$ and those gauge coupling constants are related with $g_{Y}$, the SM 
$U(1)_{Y}$ coupling constant,  by the following constraint \cite{Montero:2007cd,Montero:2016qpx}
\begin{eqnarray}
\frac{1}{g^{2}_{Y}}= \frac{1}{g^{2}_{Y^{\prime}}}+ \frac{1}{g^{2}_{BL}},
\label{runcouplingconstant}
\end{eqnarray}
where $g_{Y}$ is the Standard Model $U(1)_{Y}$ coupling constant and 
it can be written as
\begin{eqnarray}
g&=&\sqrt{\frac{8G_{F}M^{2}_{W}}{\sqrt{2}}}=0.653 \nonumber \\
g_{Y}&=&\frac{gM_{Z}}{M_{W}}
\sqrt{1- \left( \frac{M_{W}}{M_{Z}} \right)^{2}}=0.350.
\end{eqnarray}
The values for $g_{Y^{\prime}}$ is given at Fig.(\ref{fig0}).

The second and third terms, defined at Eq.(\ref{leptontotal}), are 
written as
\begin{eqnarray} 
{\cal L}^{N_{1}}&=&
\hat{\bar{N}}_{1R} e^{2[g_{Y^{\prime}} \left( \frac{5}{2} \right)
	\hat{b}_{Y^{\prime}}+g_{BL} \left( \frac{-5}{2} \right) \hat{b}_{BL}]} \hat{N}_{1R}, \nonumber \\
{\cal L}^{N_{\beta}}&=&
\sum_{\beta =2}^{3}\hat{\bar{N}}_{\beta R} e^{2[g_{Y^{\prime}} \left( \frac{-4}{2} \right)
	\hat{b}_{Y^{\prime}}+g_{BL} \left( \frac{4}{2} \right) \hat{b}_{BL}]} \hat{N}_{\beta R}\, . \nonumber \\
\label{lagrangianalepton} 
\end{eqnarray}
Therefore, our right handed neutrinos, and the right-handed sneutrinos 
are ``Weakly Sterile Neutrinos", for more detail see 
\cite{Volkas:2001zb}, and they can be Dark Matter candidate. We can 
get the same conclusion for our right-handed sneutrinos.

The second term ${\cal L}_{scalar}$, given at Eq.(\ref{l2}), is rewritten in following way
\begin{eqnarray}
{\cal L}_{scalar} &=&\left[ \int d^{4}\theta\;\left( {\cal L}^{H_{1},H_{2}}+{\cal L}^{\Phi}+{\cal L}^{\varphi} \right) \right] + 
\left(\int d^2\theta\, W+ hc\right),
\label{lagrangianascalars} 
\end{eqnarray}
where we have defined
\begin{eqnarray}
{\cal L}^{H_{1},H_{2}}&=&\hat{ \bar{H}}_{1}e^{2[g\hat{W}+g_{Y^{\prime}} \left( \frac{1}{2} \right) \hat{b}_{Y^{\prime}}]}\hat{H}_{1}+ 
\hat{ \bar{H}}_{2}e^{2[g\hat{W}+g_{Y^{\prime}} \left( \frac{-1}{2}
\right) \hat{b}_{Y^{\prime}}]}\hat{H}_{2},  \nonumber \\
{\cal L}^{\Phi} &=& \hat{ \bar{\Phi}}_{1}e^{2[g \hat{W}+g_{Y^{\prime}} \left(
\frac{5}{2} \right) \hat{b}_{Y^{\prime}}+g_{BL} \left( \frac{-6}{2}
\right) \hat{b}_{BL}]}\hat{\Phi}_{1}+ 
\hat{ \bar{\Phi}}^{\prime}_{1}e^{2[g\hat{W}+g_{Y^{\prime}} \left( \frac{-5}{2}
\right) \hat{b}_{Y^{\prime}}+g_{BL} \left( \frac{6}{2} \right)
\hat{b}_{BL}]}\hat{\Phi}^{\prime}_{1}  \nonumber \\
&+&  
\hat{ \bar{\Phi}}_{2}e^{2[g \hat{W}+g_{Y^{\prime}} \left(
\frac{-4}{2} \right) \hat{b}_{Y^{\prime}}+g_{BL} \left( \frac{3}{2}
\right) \hat{b}_{BL}]}\hat{\Phi}_{2}+ 
\hat{ \bar{\Phi}}^{\prime}_{2}e^{2[g\hat{W}+g_{Y^{\prime}} \left( \frac{4}{2}
\right) \hat{b}_{Y^{\prime}}+g_{BL} \left( \frac{-3}{2} \right)
\hat{b}_{BL}]}\hat{\Phi}^{\prime}_{2},  \nonumber \\
{\cal L}^{\varphi} &=& \hat{ \bar{\varphi}}_{1}e^{2[g_{Y^{\prime}} \left(
\frac{8}{2} \right) \hat{b}_{Y^{\prime}}+g_{BL} \left( \frac{-8}{2}
\right) \hat{b}_{BL}]} \hat{\varphi}_{1}+
\hat{ \bar{\varphi}}_{2}e^{2[g_{Y^{\prime}} \left(
\frac{-10}{2} \right) \hat{b}_{Y^{\prime}}+g_{BL} \left( \frac{10}{2}
\right) \hat{b}_{BL}]} \hat{\varphi}_{2}.  
\label{l7m2}
\end{eqnarray}
$W$ is the superpotential of this model and it is given by
\begin{equation}
W= \frac{W_{2}}{2}+ \frac{W_{3}}{3},
\label{sp1m1}
\end{equation}
where
\begin{eqnarray}
W_{2}&=&
\mu_{H}\left( \hat{H}_{1}\hat{H}_{2} \right) +
\mu_{\Phi_{1}}\left( \hat{\Phi}_{1}\hat{\Phi}^{\prime}_{1} \right) +
\mu_{\Phi_{2}}\left( \hat{\Phi}_{2}\hat{\Phi}^{\prime}_{2} \right),
\label{sp2m2}
\end{eqnarray}
Remember $\left( \hat{H}_{1}\hat{H}_{2} \right) \equiv \epsilon_{\alpha \beta}H^{\alpha}_{1}H^{\beta}_{2}$ is the usual doublet 
contraction as done in the MSSM. The free parameter $\mu_{H},\mu_{\Phi_{1}},\mu_{\Phi_{2}}$ are, in general, complex numbers and we can have CP violation interactions \cite{drees,tata}. 

The case of three chiral superfields the terms are\footnote{The terms with quarks were omitted but they are the 
same ones given at \cite{Montero:2016qpx}.}
\begin{eqnarray} 
W_{3}&=& \sum_{i=1}^{3} \left[
\sum_{j=1}^{3}f^{l}_{ij}\left( \hat{H}_{2}\hat{L}_{iL} \right) \hat{E}_{jR}+
f^{\nu}_{i}\left( \hat{\Phi}^{\prime}_{1}\hat{L}_{iL} \right) \hat{N}_{1R}+
\sum_{\beta =2}^{3}f^{\nu}_{i \beta}\left( \hat{\Phi}^{\prime}_{2}\hat{L}_{iL} \right) \hat{N}_{\beta R} \right] 
\nonumber \\ &+&
f^{N}\hat{\varphi}_{2}\hat{N}_{1R}\hat{N}_{1R}+ 
\sum_{\alpha , \beta =2}^{3}f^{N}_{\alpha \beta}\hat{\varphi}_{1}\hat{N}_{\alpha R}\hat{N}_{\beta R}. \nonumber \\
\label{sp3m2}
\end{eqnarray}
In general all the Yukawa terms defined above are complex numbers; they are symmetric in $ij$ exchange and they are 
dimensionless \cite{drees,tata}.

We see that the neutrinos $N_{1}$ and $N_{\beta}$, their $(B-L)$ quantum number are not equal, therefore 
we can not introduce the following term
\begin{eqnarray}
\hat{\varphi}_{3}\hat{N}_{1R}\hat{N}_{\beta R},
\label{mixingneutrinos}
\end{eqnarray}
because it will violate our $R$-Parity introduced 
at Sec.(\ref{Rparitym2}).

The superpotential, defined at Eq.(\ref{sp1m1}), forbid the term like $\hat{u}\hat{d}\hat{d}$ it is good because they generate the following
processes, at tree-level, that contributes to the nucleon instability:
\begin{enumerate}
\item proton decay;
\item neutron-antineutron oscillation.
\end{enumerate}

\subsection{Soft Terms to Break Supersymmetry}
\label{sec:softterms}

The soft terms, as defined at \cite{Rodriguez:2016esw,Rodriguez:2020,Rodriguez:2019,drees,tata}, are given by:
\begin{eqnarray}
{\cal L}_{Soft} &=& {\cal L}_{GMT}+ {\cal L}_{SMT} + {\cal L}_{Int}, \nonumber \\
\label{SoftSUSYm1}
\end{eqnarray}
where the term ${\cal L}_{GMT}$, known as gaugino mass term, is identical as presented at \cite{Montero:2016qpx}.

The term ${\cal L}_{SMT}$, known as scalars mass term, is given by:
\begin{eqnarray}
{\cal L}_{SMT}&=&- \left( \sum_{i=1}^{3} \left[ 
M_{L}^{2}|\tilde{L}_{i}|^{2}+
M^{2}_{l}|\tilde{E}_{i}|^{2} \right] +
M^{2}_{N_{1}}|\tilde{N}_{1}|^{2}+
\sum_{\beta =2}^{3}M^{2}_{N_{\beta}}|\tilde{N}_{\beta}|^{2}  
\right. \nonumber \\ &+& \left.
M^{2}_{H_{1}}|H_{1}|^{2}+ M^{2}_{H_{2}}|H_{2}|^{2}+ 
M^{2}_{\Phi_{1}}|\Phi_{1}|^{2}+M^{2}_{\Phi^{\prime}_{1}}|\Phi^{\prime}_{1}|^{2}+ 
M^{2}_{\Phi_{2}}|\Phi_{2}|^{2}
\right. \nonumber \\ &+& \left.
M^{2}_{\Phi^{\prime}_{2}}|\Phi^{\prime}_{2}|^{2}+
M^{2}_{\varphi_{1}}|\varphi_{1}|^{2}+M^{2}_{\varphi_{2}}|\varphi_{2}|^{2} \right) + 
\left[ \beta_{H}\left( H_{1}H_{2} \right)+
\beta_{\Phi_{1}}\left( \Phi_{1}\Phi^{\prime}_{1} \right) 
\right. \nonumber \\ &+& \left.
\beta_{\Phi_{2}}\left( \Phi_{2}\Phi^{\prime}_{2} \right) + hc \right]  , \nonumber \\
\label{smtsoft}
\end{eqnarray}
where $|H_{1}|^{2} \equiv H^{\dagger}_{1}H_{1}$ and we can redifine
\begin{eqnarray}
\beta_{H}=B_{H}\mu_{H}, \,\
\beta_{\Phi_{1}}=B_{\Phi_{1}}\mu_{\Phi_{1}}, \,\
\beta_{\Phi_{2}}=B_{\Phi_{2}}\mu_{\Phi_{2}}. 
\label{redefinirintdoiscamposescalares}
\end{eqnarray}
Notice that in the last line at equation above has the same term as introduced at 
Eq.(\ref{sp2m2}) with the the chiral superfield replaced by their scalars.

Before we continue, it is useful to remeber that the scalar mass terms
$M_{L}^{2}$ and $M^{2}_{l}$ are in general hermitean $3\!\times\!3$ matrices in generation space \cite{drees,tata}. It is very well known that the 
SUSY flavor problem occurs because the transformation that diagonalizes the fermion mass matrix does not simultaneously diagonalize the 
corresponding sfermion mass squared matrices.

The last term defined at Eq.(\ref{SoftSUSYm1}), ${\cal L}_{Int}$ has the same term as introduced at Eq.(\ref{sp3m2}) with the the chiral superfield 
replaced by theis scalars, is written as
\begin{eqnarray}
{\cal L}_{Int}&=&\left\{
\sum_{i=1}^{3}\left[ \sum_{j=1}^{3}
A^{l}_{ij}f^{l}_{ij}\left( H_{2}\tilde{L}_{iL}\right) \tilde{E}_{jR}+
A^{\nu}_{i1}f^{\nu}_{i}\left( \Phi^{\prime}_{1}\tilde{L}_{iL}\right) \tilde{N}_{1R}
\right. \right. \nonumber \\ &+& \left. \left.
\sum_{\beta =2}^{3}A^{\nu}_{i \beta}f^{\nu}_{i \beta}\left( \Phi^{\prime}_{2}\tilde{L}_{iL}\right) \tilde{N}_{\beta R} \right] 
+ A^{M}_{11}f^{N}\varphi_{2}\tilde{N}_{1}\tilde{N}_{1}
\right. \nonumber \\ &+& \left. 
\sum_{\alpha =2}^{3}\sum_{\beta =2}^{3}A^{M}_{\alpha \beta}f^{N}_{\alpha \beta}\varphi_{1}\tilde{N}_{\alpha}\tilde{N}_{\beta}+ hc  \right\}.
\label{lsoftint} 
\end{eqnarray}
The $A$-terms are known to play an important role in Affleck-Dine 
baryogenesis~\cite{Rodriguez:2020,Rodriguez:2019}, as well as in the 
inflation models based on supersymmetry~\cite{curvaton,AEGM,DHL}.

Note that, due to the $(B-L)$ invariance, the bilinear coupling proportional to $B_{M}$, see 
Eq.(\ref{egipciobl}), is not allowed. It may be generated only after the $(B-L)$ symmetry 
breaking. In this case, $B_{M}^2$ is given by 
$B_{M}^{2} \propto A_{M} u_{2}$. Therefore, as shown at \cite{Khalil:2009tm,Kajiyama:2009ae}, we can 
get one physical CP violating phase at sneutrinos mass matrix.

Before we continue, it is useful to remeber that the scalar mass terms
$M_{Q}^{2}$, $M_{u}^{2}$, $M_{d}^{2}$,
$M_{L}^{2}$, $M^{2}_{n}$ and $M_{l}^{2}$ are in general hermitean
$3\!\times\!3$ matrices in generation space \cite{tata}. It is very known that the SUSY flavor 
problem occurs because the transformation that diagonalizes the fermion mass matrix does not 
simultaneously diagonalize the corresponding sfermion mass squared matrices. In this case we can 
induce a lot of process that violate the flavor symmetry.

\section{Fermion Masses}
\label{sec:chargedfermionsmasses}

We will now discuss how to generate masses to all fermions, we will 
not to present the quarks, this case can be found at \cite{Montero:2016qpx}.

\subsection{Charged Lepton Masses}
\label{sec:leptonsmasses}

The masses of the charged leptons is given by:
\begin{eqnarray}
f^{l}_{ij}h^{0}_{2}l_{i}E_{j}+hc.
\label{chargedfermionsmasses}
\end{eqnarray}
This expression is using Weyl-spinors. We are going to define the following Dirac four components spinors
\begin{eqnarray}
{\cal E}_{a}  &=& \left(  
\begin{array}{c}  
l_{a} \\
\bar{E}_{a}     
\end{array}  \right) ,  
\label{diraceletron}
\end{eqnarray}
where $a=e, \mu , \tau$ means the physical eigenstates while $i=1,2,3$ are the symmetry eigenstates and thei are 
relationed by
\begin{eqnarray}
{\cal L}_{aL}&=&(V^{l}_{L})_{ai}l_{iL}, \nonumber \\
{\cal E}_{aR}&=&(V^{l}_{R})_{ai}E_{iR},
\label{chargedleptoneigenvectors}
\end{eqnarray} 
Using the spinors defined at Eq.(\ref{diraceletron}) the mass term become
\begin{equation}
 M^{l}_{d}\bar{{\cal E}}_{aR}{\cal L}_{aL}+hc.
\end{equation}
where $M^{l}_{d}=diag(m_{e},m_{\mu},m_{\tau})$ is the diagonal mass matrix of the charged leptons.

\subsection{Neutrino masses}
\label{sec:neutrinomasses}

The terms contributing to the neutrino masses are:
\begin{eqnarray}
- \left[
f^{\nu}_{i}\left( \Phi^{\prime}_{1}L_{iL} \right)N_{1R}+
f^{\nu}_{i \beta}\left( \Phi^{\prime}_{2}L_{iL} \right)N_{\beta R}+
f^{N}\varphi_{2}N_{1R}N_{1R}+ 
f^{N}_{\alpha \beta}\varphi_{1}N_{\alpha R}N_{\beta R}+hc 
\right].
\label{neutrinosmasses}
\end{eqnarray}
The  first two  terms  give Dirac masses term, while the last two terms  give Majorana masses terms to the neutrinos. 

Using the base 
\begin{equation}
\Psi^{0}= \left(
\begin{array}{cccccc}
\nu_{1L} & \nu_{2L} & \nu_{3L} & N_{1R} & N_{2R} & N_{3R}
\end{array}
\right)^{T} \,\ ,
\end{equation}
we can write the mass matrix to neutrino as
\begin{equation}
\left(
\begin{array}{cc}
0_{3 \times 3} & M_{D} \\
M^{T}_{D} & M_{M}
\end{array}
\right),
\end{equation}
the mass matrix leads to the 
following mass, for light $M^{\nu}_{L}$ and heavy $M^{\nu}_{P}$, neutrinos we can write the following expressions
\begin{equation}
M^{\nu}_{L}\approx - \left( M_{D} \right)^{T} \left( M_{M} \right)^{-1}M_{D}, \quad
M^{\nu}_{P}\approx M_{M},
\label{eq:seesawtoneutrinos}
\end{equation}
and we see that it can generate the type I seesaw mechanism. The 
physical states are defined as
\begin{eqnarray}
{\cal N}_{aL}&=&(V^{\nu}_{L})_{ai}\Psi^{0}_{i}, \nonumber \\
{\cal N}_{aP}&=&(V^{\nu}_{P})_{ai}\Psi^{0}_{i},
\label{neutralleptoneigenvectors}
\end{eqnarray}

Using angle $\Xi$ defined as 
\begin{eqnarray}
\tan \Xi &=& \frac{u^{\prime}_{2}}{u^{\prime}_{1}}, 
\label{angles}
\end{eqnarray}
we can write $M_{D}$ in the following way
\begin{equation}
M_{D}= \frac{u^{\prime}_{1}}{2 \sqrt{2}}\left(
\begin{array}{ccc}
f^{\nu}_{1} & f^{\nu}_{12} \tan \Xi & f^{\nu}_{13} \tan \Xi \\
f^{\nu}_{2} & f^{\nu}_{22} \tan \Xi & f^{\nu}_{23} \tan \Xi \\
f^{\nu}_{3} & f^{\nu}_{32} \tan \Xi & f^{\nu}_{33} \tan \Xi
\end{array}
\right),
\label{diracneutrino}
\end{equation}
if $\Xi = ( \pi /2)$rad our matrix is identical to the $M_{D}$ presented at Ref.~\cite{Montero:2011jk}.

In order to write $M_{M}$, first we have to defined $\Gamma , \Omega$ though the following ratios
\begin{eqnarray}
\tan \Gamma &=& \frac{w_{1}}{u^{\prime}_{1}}, \nonumber \\
\tan \Omega &=& \frac{w_{2}}{u^{\prime}_{1}},
\label{angles1}
\end{eqnarray}
using these new parameters we get
\begin{equation}
M_{M}= \frac{u^{\prime}_{1}}{2 \sqrt{2}}\left(
\begin{array}{ccc}
f^{N} \tan \Omega & 0 & 0  \\
0 & f^{N}_{22} \tan \Gamma & f^{N}_{23} \tan \Gamma \\
0 & f^{N}_{32}\tan \Gamma & f^{N}_{33} \tan \Gamma
\end{array}
\right),
\label{majorananeutrino}
\end{equation}
our matrix is identical to the $M_{M}$ presented at Ref.~\cite{Montero:2011jk}.

\subsection{Leptonic Mixing}
\label{sec:neutrinomasses}

The Yukawa terms came from our superpotential, defined at 
Eq.(\ref{sp3m2}). Above we discuss the mass expressions, Now we want to 
discuss the mixing parameters, The lepton mass matrices take the 
diagonal form when we perform the unitary transformations defined 
at Eqs.(\ref{chargedleptoneigenvectors},\ref{neutralleptoneigenvectors}). 
In this base, we get the usual leptonic charged current interaction 
\begin{equation}
- \frac{g}{\sqrt{2}}\left( \bar{{\cal L}}_{iL}\gamma^{m} 
\left( V^{l \dagger}_{L}V^{\nu}_{L} \right)_{ij}{\cal N}_{jL}W_{m}
+hc \right),
\end{equation}
where the lepton flavour mixing matrix is then given by
\begin{equation}
U_{PMNS}= V^{l \dagger}_{L}V^{\nu}_{L}, 
\end{equation}
this lepton mixing matrix is known as 
Pontecorvo-Maki-Nakagawa-Sakata (PMNS) matrix. We can conclude our model is also compatible with the observed solar and atmospheric mass scales and the tribimaximal 
mixing matrix, for more details see Ref.~\cite{Montero:2011jk}, their Sec.IV.

In this class of models the sfermion $\tilde{f}$ mass matrix is 
$12 \times 12$ hermitian matrix, and it is given by
\begin{equation}
{\cal M}^{2}_{\tilde{f}}= \left(
\begin{array}{cc}
{\cal M}^{2}_{\tilde{f}_{L}\tilde{f}_{L}} & 
{\cal M}^{2}_{\tilde{f}_{L}\tilde{f}_{R}} \\
{\cal M}^{2}_{\tilde{f}_{R}\tilde{f}_{L}} & 
{\cal M}^{2}_{\tilde{f}_{R}\tilde{f}_{R}}
\end{array}
\right)
\end{equation}
and it is known as super-CKM and super-PMNS basis. 

To simplify the calculation we further assume
that all diagonal entries in 
${\cal M}^{2}_{\tilde{f}_{L}\tilde{f}_{L}}$,
${\cal M}^{2}_{\tilde{f}_{L}\tilde{f}_{R}}$, 
${\cal M}^{2}_{\tilde{f}_{R}\tilde{f}_{L}}$ and
${\cal M}^{2}_{\tilde{f}_{R}\tilde{f}_{R}}$ 
are set to be equal to the common value
$M^{2}_{\rm{SUSY}}$, and then we normalize the off-diagonal elements
to $M^{2}_{\rm{SUSY}}$ \cite{drees,tata},
\begin{eqnarray}
&& (\delta^{\tilde{f}}_{LL})^{ij} =
\frac{({\cal M}^{2})^{ij}_{\tilde{f}_{L}\tilde{f}_{L}}}{M^2_{\rm{SUSY}}}\,,
\hspace{1.0truecm} 
(\delta^{\tilde{f}}_{RR})^{ij} =
\frac{({\cal M}^{2})^{ij}_{\tilde{f}_{R}\tilde{f}_{R}}}{M^2_{\rm{SUSY}}}\,,
\hspace{1.0truecm} \nonumber \\
&& (\delta^{\tilde{f}}_{LR})^{ij} =
\frac{({\cal M}^{2})^{ij}_{\tilde{f}_{L}\tilde{f}_{R}}}{M^2_{\rm{SUSY}}}\,,
\hspace{1.0truecm} 
(\delta^{\tilde{f}}_{RL})^{ij} =
\frac{({\cal M}^{2})^{ij}_{\tilde{f}_{R}\tilde{f}_{L}}}{M^2_{\rm{SUSY}}}\,,
\hspace{1.0truecm} (i \ne j,i,j=1,2,3). \nonumber \\ \label{deltadefb}
\end{eqnarray}
and we want to analyse it in a further work. 

Indeed, masses and mixings of sparticles are of crucial importance both
theoretically and experimentally\cite{drees,tata}: 
\begin{itemize}
\item[1-] they determine the properties of the sparticles searched;
\item[2-] they are directly related to the question of how SUSY is broken. 
\end{itemize}

\begin{figure}[t]
\begin{center}
\epsfig{file=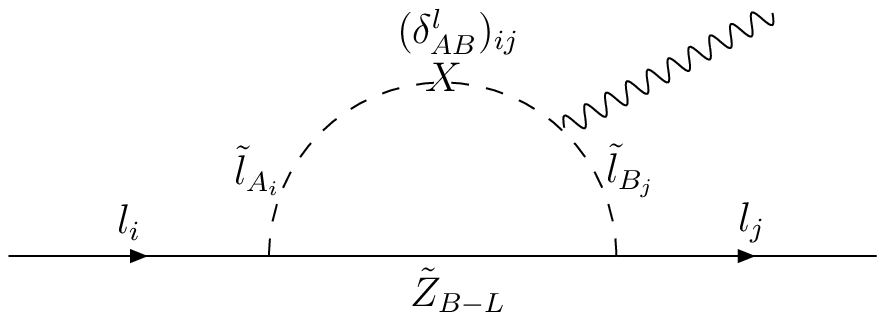, width=7.5cm,height=3.5cm,angle=0}
\epsfig{file=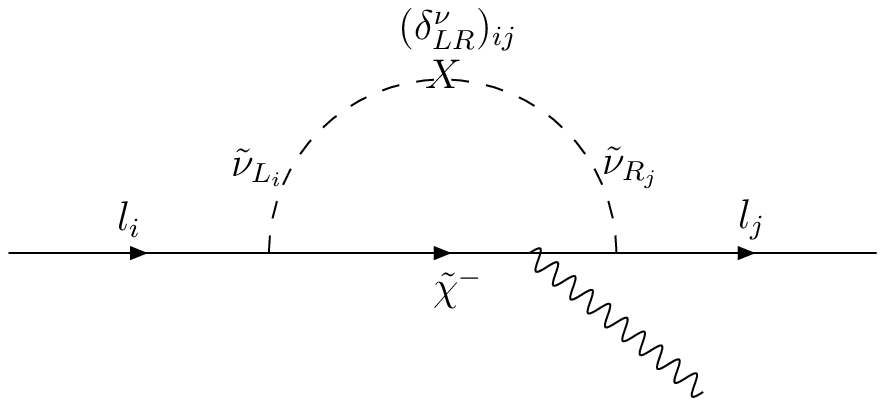, width=7.5cm,height=3.5cm,angle=0}
\end{center}
\vspace{-0.5cm} \caption{New contributions to the decay 
$l_{i} \to l_{j} \gamma$ in SUSY $(B-L)$ model taken from \cite{Khalil:2009tm}, for $\delta$ see 
Eq.(\ref{deltadefb}).} 
\label{fig:3}
\end{figure}

\begin{figure}[t]
\begin{center}
\epsfig{file=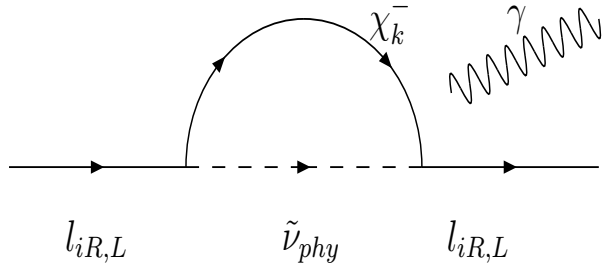,width=8cm,height=3.5cm}
\caption{The chargino contributions to the charged lepton EDM due
to chargino and sneutrino exchange taken from \cite{Kajiyama:2009ae}.}
\label{edmfig}
\end{center}
\end{figure}

The lack of observation  of the decays $\mu \rightarrow e \gamma$, $\tau \rightarrow \mu \gamma$
and $\tau \rightarrow e \gamma$ put some constraints on the lepton-slepton coupling. In the case 
of $(B-L)$ supersymmetric model has an additional contribution shown at Fig.(\ref{fig:3}) as 
discussed at \cite{Khalil:2009tm}. However, the most stringent
constraints on the SUSY phases come from continued efforts to
measure the electric dipole moments (EDM) of the neutron and in the case of $(B-L)$ supersymmetric 
model it was studied at \cite{Kajiyama:2009ae}, where we take Fig.(\ref{edmfig}).

By another hand, processes like $b \rightarrow s \gamma$ decay and the measurements of mass
difference in $B^{0} \bar{B}^{0}$ and
$D^{0} \bar{D}^{0}$ mass differences  yield
constraints on the quark-squark coupling, the most stringent restrictions here come
from what is already known about $K^{0} \bar{K}^{0}$ mixing.

\section{Leptogenesis}
\label{sec:leptom1}

Leptogenesis \cite{Pelto:2010vq} is a term for a scenario where new physics generates a
lepton asymmetry in the Universe which is partially converted to a
baryon asymmetry. In the review \cite{vietnam}
we learned that the introduction of singlet neutrinos with
Majorana masses and Yukawa couplings to the doublet leptons fulfills
Sakharov conditions. Then we can expected in explain the matter asymmetry with 
this new model as done in \cite{Pelto:2010vq,Iso:2010mv}.

The mechanisms to create a baryon asymmetry from an initially symmetric state must in general satisfy the 
three basic conditions for baryogenesis as pointed out by Sakharov in 1967
\footnote{They are known as Sakharov conditions} \cite{sakharov}: 
\begin{enumerate}
 \item Violate baryon number, $B$, conservation;
 \item Violate $C$ and $CP$, conservation and
 \item to be out of thermal equilibrium.
\end{enumerate}
It is found that the $CP$ violation observed in the quark sector \cite{KM_CPviolation} (e.g. in 
$K^0$-$\bar{K}^0$ or $B^0$-$\bar{B}^0$ mesons system) is far too small \cite{CP_in_SM_tooSmall} 
to give rise to the observed baryon asymmetry, therefore these conditions are extended to include 
lepton number ($L$) violation processes.

\begin{figure}[tb]
\begin{center}
 \includegraphics[width=\textwidth]{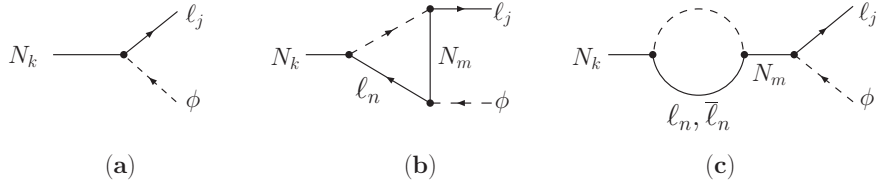}
\end{center} 
\caption{The {\bf (a)} tree-level,  {\bf (b)} one-loop vertex correction,  and {\bf (c)} one-loop self-energy correction graphs for the decay: 
$N_{k} \rightarrow l_{j}\,\overline{\phi}$. This figure was taken 
from \cite{Law:2009vh}.}
\label{fig1:lepto_std_decay_graphs}
\end{figure}

All Sakharov's conditions for leptogenesis will be satisfied if these decays also violate $CP$ 
and go out of equilibrium at some stage during the evolution of the early universe. 
The requirement for $CP$ violation means that the coupling matrix $Y$ must be complex and the 
mass of $N_k$ must be greater than the combined mass of $l_{j}$ and $\phi$, so that interferences 
between the tree-level process (Fig.~\ref{fig1:lepto_std_decay_graphs}a) and the one-loop 
corrections (Fig.~\ref{fig1:lepto_std_decay_graphs}b, c) with on-shell intermediate states will 
be nonzero \cite{Law:2009vh,Law:2010zz}.

In our model the massive neutrinos can decay thought the following vertices in the tree-level
\begin{itemize}
\item[1] $f^{\nu}_{i}$
\begin{itemize}
\item[a] $N_{1R}\rightarrow \nu_{iL}\phi^{\prime 0}_{1}$ \,\ 
$\tilde{N}_{1R}\rightarrow \nu_{iL}\tilde{\phi}^{\prime 0}_{1}$
\item[b] $N_{1R}\rightarrow l_{iL}\phi^{\prime +}_{1}$ \,\
$\tilde{N}_{1R}\rightarrow l_{iL}\tilde{\phi}^{\prime +}_{1}$
\end{itemize}
\item[2] $f^{\nu}_{i \beta}$
\begin{itemize}
\item[a] 
$N_{\beta R}\rightarrow \nu_{iL}\phi^{\prime 0}_{1}$ \,\ 
$\tilde{N}_{\beta R}\rightarrow \nu_{iL}\tilde{\phi}^{\prime 0}_{1}$
\item[b] 
$N_{\beta R}\rightarrow l_{iL}\phi^{\prime +}_{1}$ \,\ 
$\tilde{N}_{\beta R}\rightarrow l_{iL}\tilde{\phi}^{\prime +}_{1}$
\end{itemize}
\end{itemize}
Therefore, we have all necessary condition to generate viabel Leptogenesis in this model, in a 
similar way as shown at \cite{Kajiyama:2009ae}. We expect we can generate one viavel Leptogenesis, 
viavel CP violation processes and also an invisibel axion in this model.  

\begin{figure}[ht]
\begin{center}
\vglue -0.009cm
\mbox{\epsfig{file=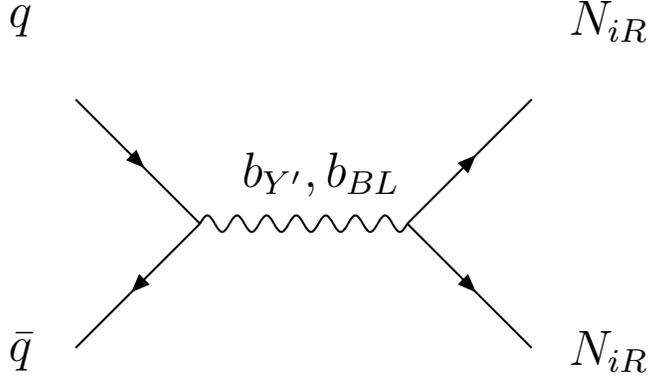,width=0.7\textwidth,angle=0}}
\end{center}
\caption{Process $\bar{q}q \rightarrow N_{iR} N_{iR}$.}
\label{figqqNN}
\end{figure}

\begin{figure}[ht]
\begin{center}
\vglue -0.009cm
\mbox{\epsfig{file=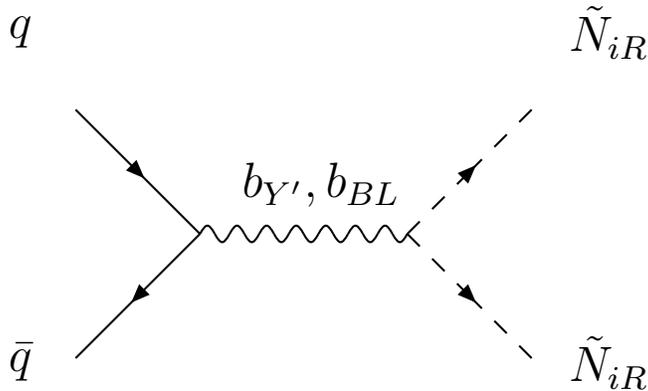,width=0.7\textwidth,angle=0}}
\end{center}
\caption{Process $\bar{q}q \rightarrow \tilde{N}_{iR} \tilde{N}_{iR}$.}
\label{figqqtNtN}
\end{figure}

We can have the following $L$ violation process
\begin{itemize}
\item[1] $\bar{q}q \rightarrow N_{iR} N_{iR}$;
\item[1] $\bar{q}q \rightarrow \tilde{N}_{iR} \tilde{N}_{iR}$. 
\end{itemize}
Their Feynman diagram is shown at Figs.(\ref{figqqNN},\ref{figqqtNtN}). 
Their experimental signal can be $l^{-}l^{+}$ plus jet due charginos 
and neutralinos decaying modes. It can be found to 
some collider like LHC.

\section{Conclusions}
\label{sec:conclusion}

We presented the Supersymmetric version of the model with three distintics right handed neutrinos presented at \cite{Montero:2007cd}. This model has some interesting facts such as we can 
generate neutrinos masses via see-saw mechanism; they preserve the $R$-parity, therefore the 
neutrinos and neutralinos are distint particles and due this fact the lightest supersymmetric
particle (LSP) is stable and the sparticles are pair produced in any collider experiment and it is 
also a good candidate for Dark Matter in Universe; we can generate a viable Leptogenesis scenario due 
the Majorana phases in the sneutrnio mass matrix that will induce the decay of Heavy Neutrinos in 
leptons plus usual scalars generating in this way more leptons than antileptons. We think that these models 
have som nice predictios that could be explored in the near future.

\end{document}